\documentstyle[aps,prl,preprint]{revtex}
\include{psfig}

\tightenlines

\begin{document}
\draft

\title{Fluctuation-dissipation relations and energy landscape in an 
out-of-equilibrium strong glass-forming liquid}

\author{Antonio Scala$^{1}$, Chantal Valeriani$^{1}$,
Francesco Sciortino$^{1,2}$ and Piero Tartaglia $^{1,2}$}

\address{$^{1}$ Dipartimento di Fisica e Istituto Nazionale per la
Fisica della Materia, \\ Universit\'a di Roma {\it La Sapienza}, P.le
Aldo Moro 2, I-00185, Roma, Italy }

\address{$^{2}$INFM, Center for Statistical Mechanics and Complexity,
Universit\'a di Roma {\it La Sapienza}
}

\date{27 Sept 2002}
\maketitle
\begin{abstract}
We study the out-of-equilibrium dynamics following a temperature-jump
in a model for a strong liquid, BKS-silica, and compare it with the
well known case of fragile liquids. We calculate the
fluctuation-dissipation relation, from which it is possible to
estimate an effective temperature $T_{eff}$ associated to the slow
out-of-equilibrium structural degrees of freedom. We find the striking
and unexplained result that, differently from the fragile liquid
cases, $T_{eff}$ is smaller than the bath temperature.
\end{abstract}
\bigskip

A central issue of statistical physics is to link the static
properties of a thermodynamical system to its dynamics. A fundamental
result in this direction has been the derivation of the
fluctuation-dissipation theorem (FDT)~\cite{fdt_storic}, which links
the response functions to autocorrelation functions; however,
derivations of the FDT require the system to be in equilibrium and the
response to be in linear regime.

Recently, a great deal of work has concentrated on the generalization
of the FDT to out-of-equilibrium (OOE) situations~\cite{fdt_new}; in
particular, it has been proposed to generalize FDT in the
form~\cite{fdt_gen}

\begin{equation}
R_{AB}(t,t')=\frac{X(C_{A,B}(t,t'))}{k_B T} \frac{\partial C_{AB}(t,t')}{
\partial t'}
\label{eq:genFDT}
\end{equation}

\noindent 
where $A$ and $B$ are conjugated observables, $R_{AB}(t,t')$ is the
mutual response of the observable $A$ to an impulsive conjugated field
applied at $t'$,
$C_{AB}(t,t')=\left<A(t)B(t')\right>-\left<A(t)\right>\left<B(t')\right>$
is the unperturbed correlation function of the observables and $T$ is
the temperature of the bath; $X$ is supposed to be a function of
$C_{AB}$ only and equals to one in equilibrium and linear regime. In
recent analytical and numerical work, it has been observed that the
out of equilibrium dynamics which follows a $T$-jump is characterized
by an FDT relation with $X=1$ at short times and $X<1$ at long
times~\cite{fdt_goodexamples}. It has been proposed to interpret the
ratio $T_{eff}=T/X$ as an effective temperature associated to the
different timescales of the system~\cite{timescales}: while fast
degrees of freedom are supposed to be thermalised so that $T_{eff}$
equals the bath temperature $T$ (and hence $X=1$ and ordinary FDT
holds), slow degrees of freedom are still reminiscent of the
temperature of the system before being brought in OOE so that $T_{eff}
\neq T$ (i.e. $X\neq 1$). In general, it is observed that for
quenching experiments $T_{eff} > T$ for the slow degrees of freedom;
also in the case of a ``reverse'' quench, where the system is suddenly
heated up, slow degrees of freedom ``remember'' the initial state
being $T_{eff}<T$~\cite{fdtXY}. For fragile liquids, the results are
consistent with the one step replica symmetry breaking (1RSB) scenario
both for quenches~\cite{fdt_goodexamples,fsT} and for
``crunches''~\cite{fdtRUOCCO} (sudden changes of volume); in this
scenario, $X$ becomes a step function equal to one at short times
($T_{eff}=T$) and equal to a constant less than one ($T_{eff}>T$) at
long times. Exceptions to this behavior are pointed out in
refs.~\cite{fdtKRYS,fdt_strong,actFDT}.

The slow dynamics of supercooled systems have been considered also
under a complimentary approach, based on the statistical properties of
the free-energy~\cite{free_land} or the potential
energy~\cite{pot_land} landscape. In the landscape approach the free
energy of a system is separated in configurational contributions,
linked to the basins of attraction of local minima of the landscape,
and in vibrational contributions linked to the number of states in
such basins. It has been suggested that fast relaxations are related
to the exploration of states inside a basin, while slow relaxations
correspond to the exploration of different potential energy landscape
basins. The topology of the landscape explored should therefore
correlate to the dynamics of the system.  Support of this
interpretation for fragile liquids has been presented both for
equilibrium~\cite{NatureLand} and out-of-equilibrium
conditions~\cite{fsLAND-OOE}.

Under the assumption that the basins explored in OOE are the same that
would be explored in equilibrium~\cite{fsAGING}, it is possible to
extend the landscape approach to OOE situations at the expenses of
adding an additional parameter to the thermodynamic formalism. In OOE,
the landscape-based liquid free energy can be written as~\cite{fsT}

\begin{equation}
f(e_{is},R) = e_{is} - T_{int} s_{conf}(e_{is}) + f_{vib}(e_{is},T)
\end{equation}

\noindent 
where $e_{is}$ is the average energy of the local potential energy
minima (the so called ``inherent structures''~\cite{pot_land}), the
configurational entropy $s_{conf}$ counts the number of minima of
energy $e_{is}$, $f_{vib}$ is the average free energy of a basin of
depth $e_{is}$, $T$ is the temperature of the bath and $T_{int}$ is a
Lagrange multiplier that extremizes $f$; in the equilibrium case,
$T_{int}$ is just the temperature of the bath $T$. The extremum
condition $\partial f /\partial e_{is}=0$ gives an expression of the
internal temperature $T_{int}$

\begin{equation}
T_{int}(e_{is},T)=
\left(1+\frac{\partial f_{vib}(e_{is},T)}{\partial e_{is}}\right) /
\frac{\partial s_{conf}(e_{is})}{\partial e_{is}}
\label{eq:Tint}
\end{equation}

\noindent
that can be calculated from the equilibrium values of $s_{conf}$ and
from analytical approximations for $f_{vib}$. In the case of a
prototype fragile liquid, the binary mixture Lennard-Jones, it has
been verified that in OOE conditions the configurational temperature
$T_{int}$ calculated from the landscape approach and the effective
temperature $T_{eff}$ measured from the extension of the FDT
coincide~\cite{fsT}. This coincidence is supported from the analytical
predictions for the prototype fragile mean field system, the p-spin
model, for which again $T_{int}=T_{eff}$~\cite{franzT}.

In the case of strong liquids, the situation is less developed.  Only
recently, a realistic model for liquid silica, the BKS
model~\cite{BKSpot}, has been studied in detail in equilibrium
conditions.  At variance with fragile systems, where the $T$
dependence of the characteristic times is always super-Arrhenius,
BKS-silica shows a high $T$ region ($T>3330 K$) where the dynamics is
super-Arrhenius and a low-$T$ region ($T<3300K$) where the equilibrium
dynamics recovers an Arrhenius behavior. For $T>3330 K$, the dynamics
of BKS silica is well described by mode-coupling theory; at difference
with other fragile liquids, the presence of local tetrahedral
structures and therefore of a strong oxygen-silicon-oxygen
correlation, makes also three-point correlation functions to be taken
in account in order to get quantitative agreement with the
theory~\cite{mctstrong}. For $T<3330 K$, BKS silica becomes strong, in
agreement with the behavior of real silica. Interestingly enough,
despite this strong-to-fragile crossover, the structural relaxation
times of BKS-silica are well described in the whole temperature region
by the Adam-Gibbs formula~\cite{fsBKS-AG} as in the case of fragile
liquids; moreover, free directions in configuration space vanish at
the mode-coupling temperature~\cite{fsBKS-INM}. Therefore, landscape
theory is able to rationalize the equilibrium dynamics of fragile and
strong systems despite the alleged diversity of their
landscapes~\cite{angellDivLand}.

It is therefore important to check if $T_{int}=T_{eff}$ also in the
case of strong liquids: if this would be the case, we could have
afforded a simple thermodynamic understanding of OOE liquids in terms
of two temperatures, the kinetic temperature $T$ associated with the
fast vibrational degrees of freedom and the internal temperature
$T_{int}$ (or its dynamical counterpart $T_{eff}$) associated to the
slowly evolving configurational degrees of freedom~\cite{note:2T}.

In our simulations we have employed the BKS-potential~\cite{BKSpot}
with the parameters of ref.~\cite{fsBKS-AG}. Simulations are for
systems of $N=999$ particles ($N_{Si}=333$ and $N_{O}=666$) at fixed
density $\rho=2.36 g/cm^3$ with a Nose'-Hoover thermostat. Long range
forces have been calculated implementing the Ewald sums.  We have
quenched $280$ independent configurations equilibrated at $T=3800 K$
(well above the fragile-to-strong crossover at $T=3330
K$~\cite{BKSkob}) at two different final temperatures $T_1=2900$ and
$T_2=2500$; both these temperatures are in the region where BKS-silica
has a strong behavior~\cite{BKSkob}. The Nose'-Hoover parameters have
been selected to bring the kinetic energy of the system in equilibrium
within 0.5 ps. We have measured two unperturbed correlation functions
of the form $ \left< A(t+t_w)B(t_w) \right>_0$, where $A$ and $B$ are
conjugated observables with zero average in absence of a
perturbation. Indicating with ${\vec r}^o_l$ the coordinates of the
$l$-th oxygen atom ($l=1..N_O, N_O=666$), the choice $A =
\rho^o_{coll} = \sum_{l}^{N_O} \exp\left( i{\vec k}\cdot {\vec
r}^o_l \right) / \sqrt{N}$ and
$B=(\rho^o_{coll}+\rho^{o*}_{coll})$ corresponds to the dynamical
structure factor $C_{coll}$; the choice $A = \rho^o_{self} =
\exp\left( i{\vec k}\dot {\vec r}^o_l \right)$ and
$B=(\rho^o_{self}+\rho^{o*}_{self})$ corresponds to the self part of
the intermediate scattering function $C_{self}$. In the case of the
self correlation, averages over several ions have been performed.  The
total number of independent realizations (different configurations,
different wave vectors, different ions) is $\sim 30000$ for the
quenches at $T_1$ and $\sim 20000$ for the quenches at $T_2$, both for
the self and for the collective case.

\begin{figure}
\begin{center}
\mbox{
    \psfig{figure=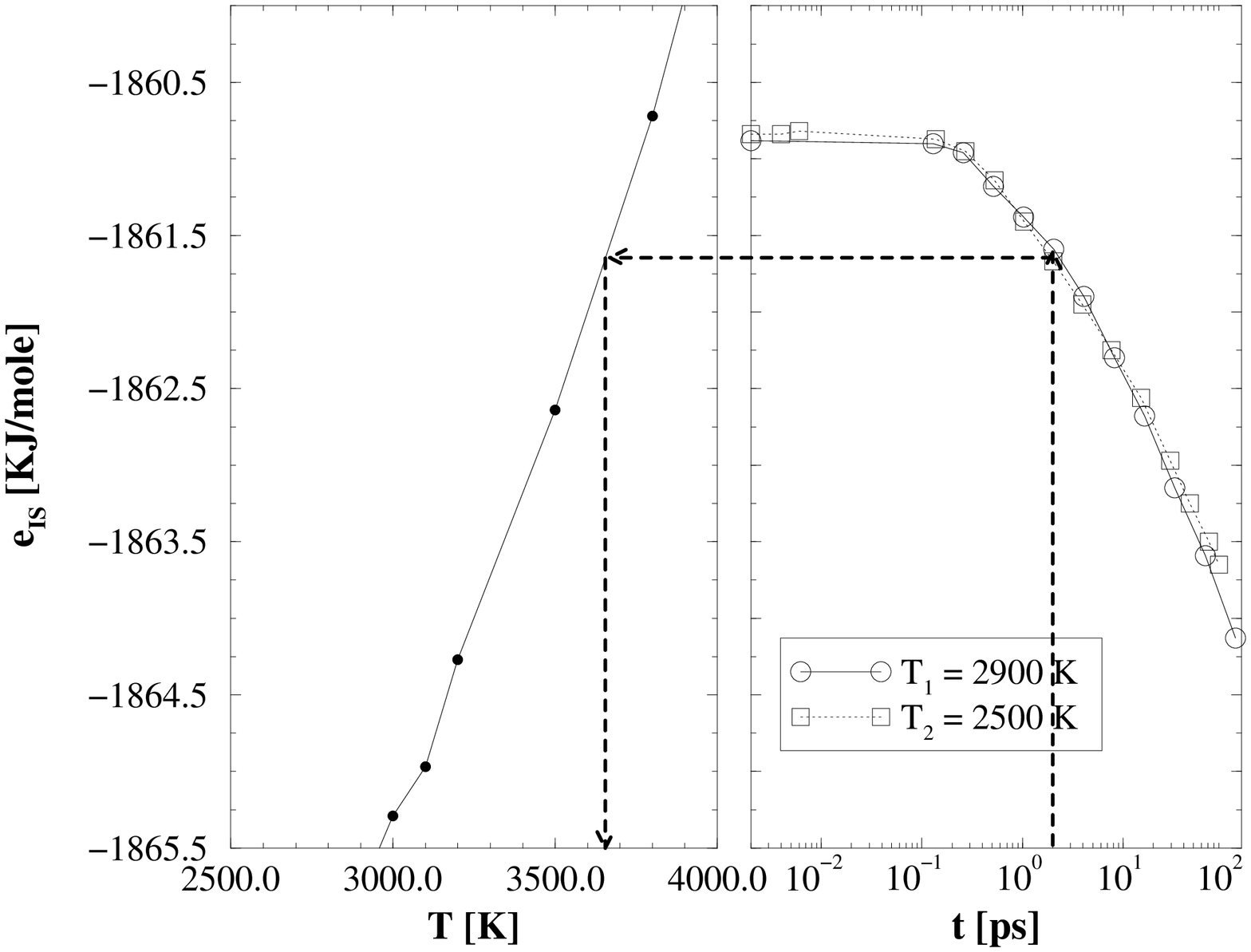,width=14.0cm,angle=-90}
}
\end{center}

\caption{
Left panel: equilibrium relation between $e_{is}$ and $T$.  Right
panel: $e_{is}$ as a function of $t$ for the two quenches
considered. The arrows show graphically the procedure that connects
the value of $e_{is}(t)$ to the equilibrium temperature $T(e_{is})$
for $t = t_w = 2ps$; landscape theory therefore predicts the
temperature associated to the configurational degrees of freedom to be
$T_{int} \approx 3650$~$K$ in both quenches.
}
\label{fig:Tint}
\end{figure}

Fig.~\ref{fig:Tint} shows the evolution of the average inherent
structure energy during the aging, following the $T$-jump and compares
it with the known equilibrium values. The average $e_{IS}$ decreases
with time, while the system searches for deeper and deeper basins in
the attempt to equilibrate.  We note that the $e_{is}$ decrease does
not depend on the bath temperatures selected, suggesting that the
system is exploring configuration space with a saddle-dominated
entropic dynamics~\cite{note:freexpl}.  For example, after 2 ps,
independently from the bath temperature, the system is exploring
basins which are typically explored in equilibrium at $T \approx 3650
K$.

By calculating the density of states of the explored basins and
comparing it with the corresponding quantity evaluated in
equilibrium, we confirm that the basins explored during the aging
dynamics have the same density of states of the basins explored in
equilibrium. Since in the case of BKS-silica, the density of states
does not depends on $e_{is}$, in harmonic approximation $\partial
f_{vib}/\partial e_{is}=0$. This implies that, according to
Eq.~\ref{eq:Tint}, the $T_{int}$ of the BKS-system when populating
basins of average depth $e_{is}$ coincides with the temperature at
which basins of depth $e_{is}$ are populated in equilibrium
(Fig.\ref{fig:Tint}).  It is therefore possible to calculate the
internal temperature of the system inverting the equilibrium relations
between $e_{is}$ and $T$~\cite{fsT} (Fig.~\ref{fig:Tint}). At the
waiting time $t_w = 2 ps$, we find that $T_{int} \approx 3650 K$ for
both the quenches, so that $T_{int} > T_1,T_2$ and we expect to
measure for the slow degrees of freedom a $T_{eff}$ higher than the
bath temperature.

\begin{figure}
\mbox{
    \psfig{figure=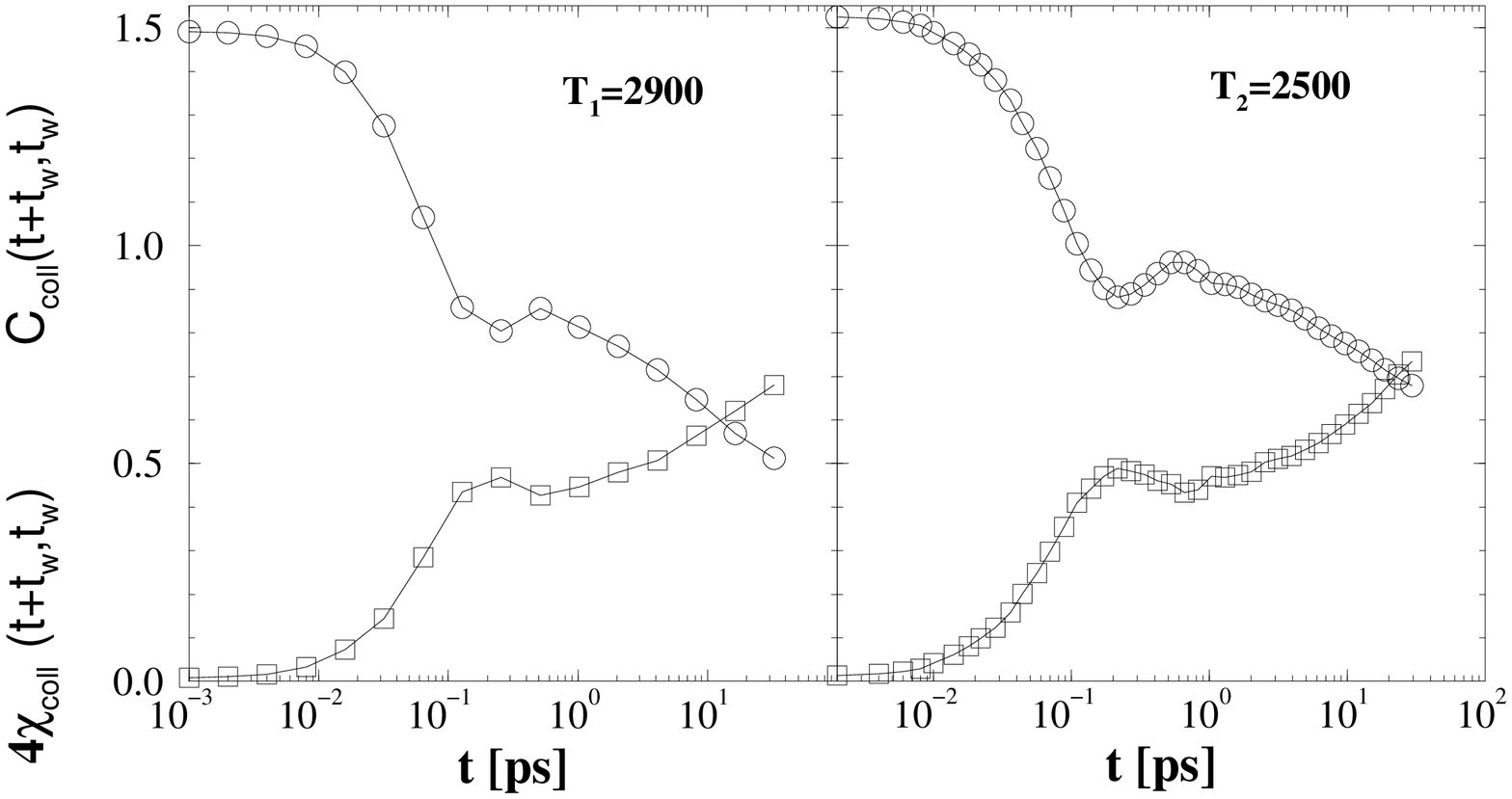,width=14.0cm,angle=-90}
}
\vspace*{0.1cm}
\mbox{
    \psfig{figure=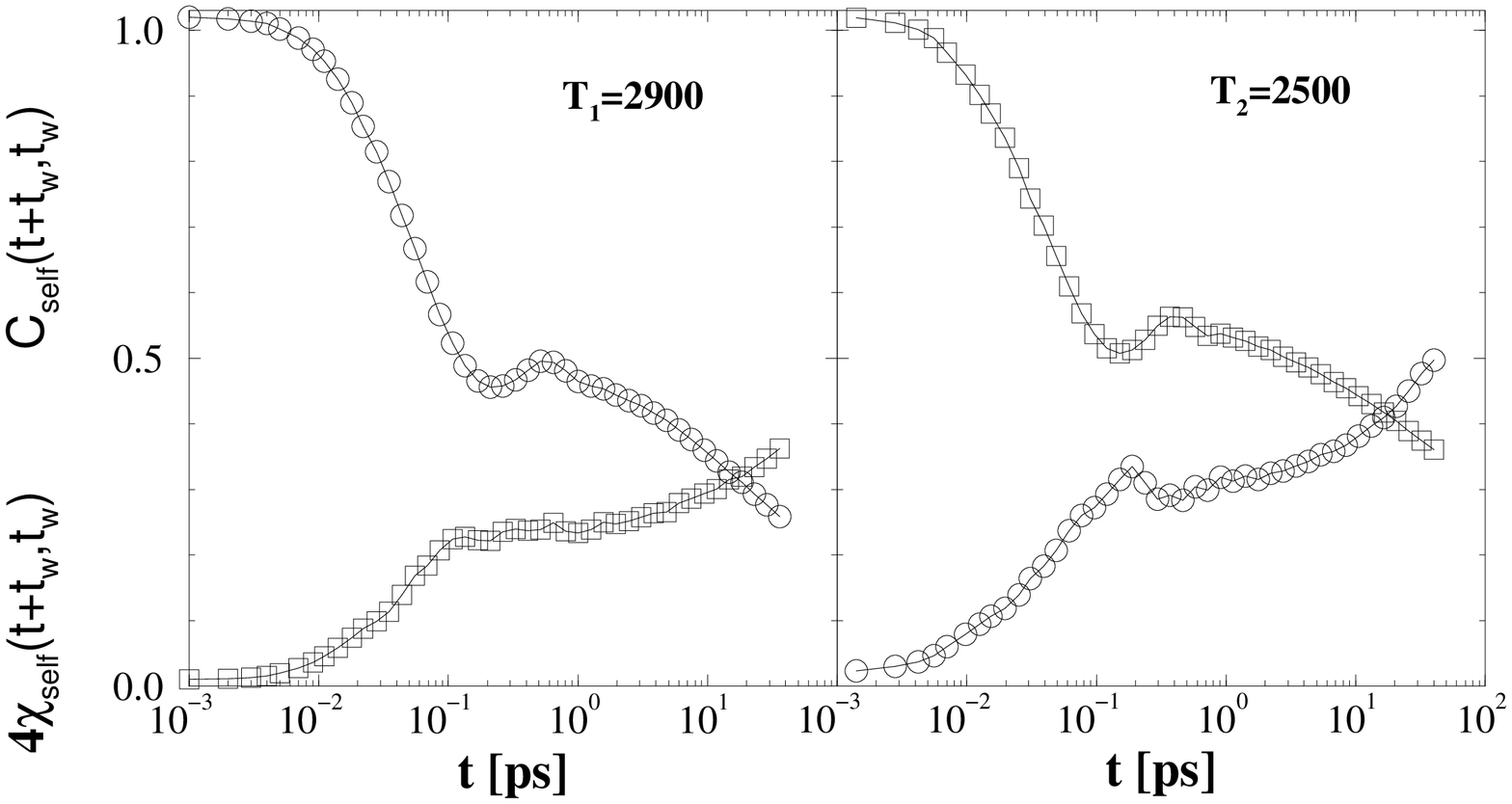,width=14.0cm,angle=-90}
}

\caption{
Correlation and response functions for the collective (upper panels)
and the self (lower panels) variables, at the two temperatures
$T_1=2900$~$K$ (left panels) and $T_2=2500$~$K$. Short time behavior
corresponds to the fast vibrational dynamics inside a basin; the
corresponding first decay of the correlation functions to a plateau is
fast. Long time behavior corresponds to the exploration of new minima
causing the slow decay of correlation functions from the plateaus.
}
\label{fig:CandR}
\end{figure}

In order to measure $T_{eff}$, we evaluate the integrated response
function $\chi=V_0\int^t_{t_w}dt'R(t,t')$; from Eq.~\ref{eq:genFDT}
the slope $\frac{d\chi}{dC}$ of the parametric plot of $\chi$ versus
$C$ yields $V_0\frac{X(C)}{k_B T}=-V_0\frac{1}{k_B T_{eff}}$. We have
applied conjugate fields $V_0 B \theta(t-t_w)$ of constant amplitude
$V_0$ after a waiting time $t_w=2ps$, significatively bigger than the
characteristic time of the microscopic dynamics (see
Fig.~\ref{fig:CandR}), and measured the integrated responses
$\chi_{coll}=\rho^{o}_{coll}$ and $\chi_{self}=\rho^o_{self}$.  We
have chosen $V_0=4.0kJ/mol$ for $B=(\rho^o_{coll}+\rho^{o*}_{coll})$
and $V_0=5.0J/mol$ for $B=(\rho^o_{self}+\rho^{o*}_{self})$ in order
to have a good signal-to-noise ratio in the response function while
being still well inside the linear regime
region~\cite{note:linreg}. We have averaged response and correlation
functions over $300$ wave vectors of length $|{\vec k}|=28.6 nm^{-1}$
corresponding to the main peak of the static structure factor in order
to have a good signal-to-noise ratio for the correlation
functions~\cite{note:stn}. We show in Fig.~\ref{fig:CandR} the
correlations and the response functions of the system.  At the $t_w$
chosen, exploration of deeper basins has started as shown in
Fig.\ref{fig:Tint}. This waiting time, which is much shorter than the
infinite-time limit in which analytical predictions are derived, is
consistent with the aging time chosen in most of previous studies of
fragile liquids~\cite{note:biggertw}.

\begin{figure}
\begin{center}
\mbox{
    \psfig{figure=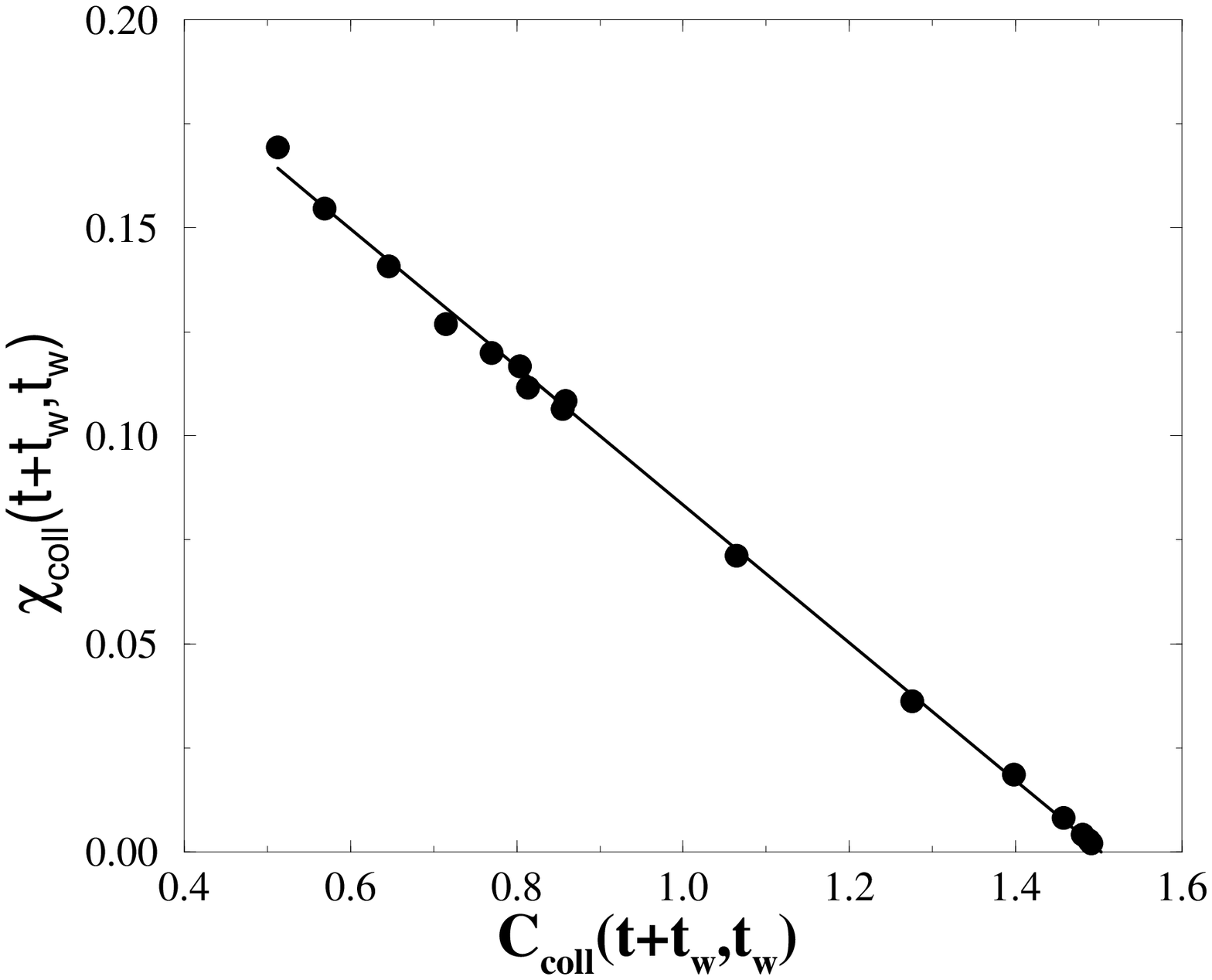,width=7.0cm,angle=-90}
    \hspace*{0.1cm}
    \psfig{figure=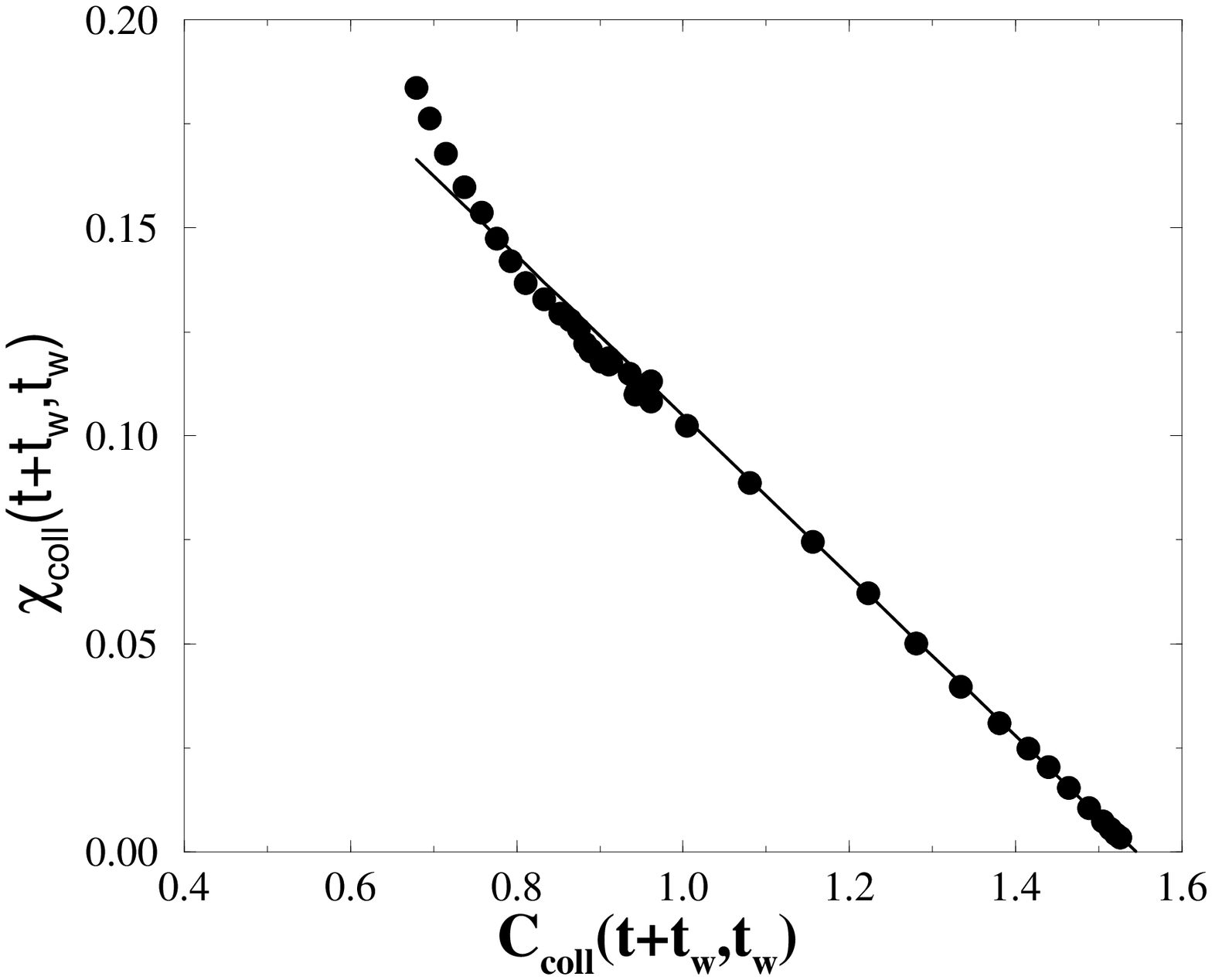,width=7.0cm,angle=-90}
}
\vspace*{0.1cm}
\mbox{
    \psfig{figure=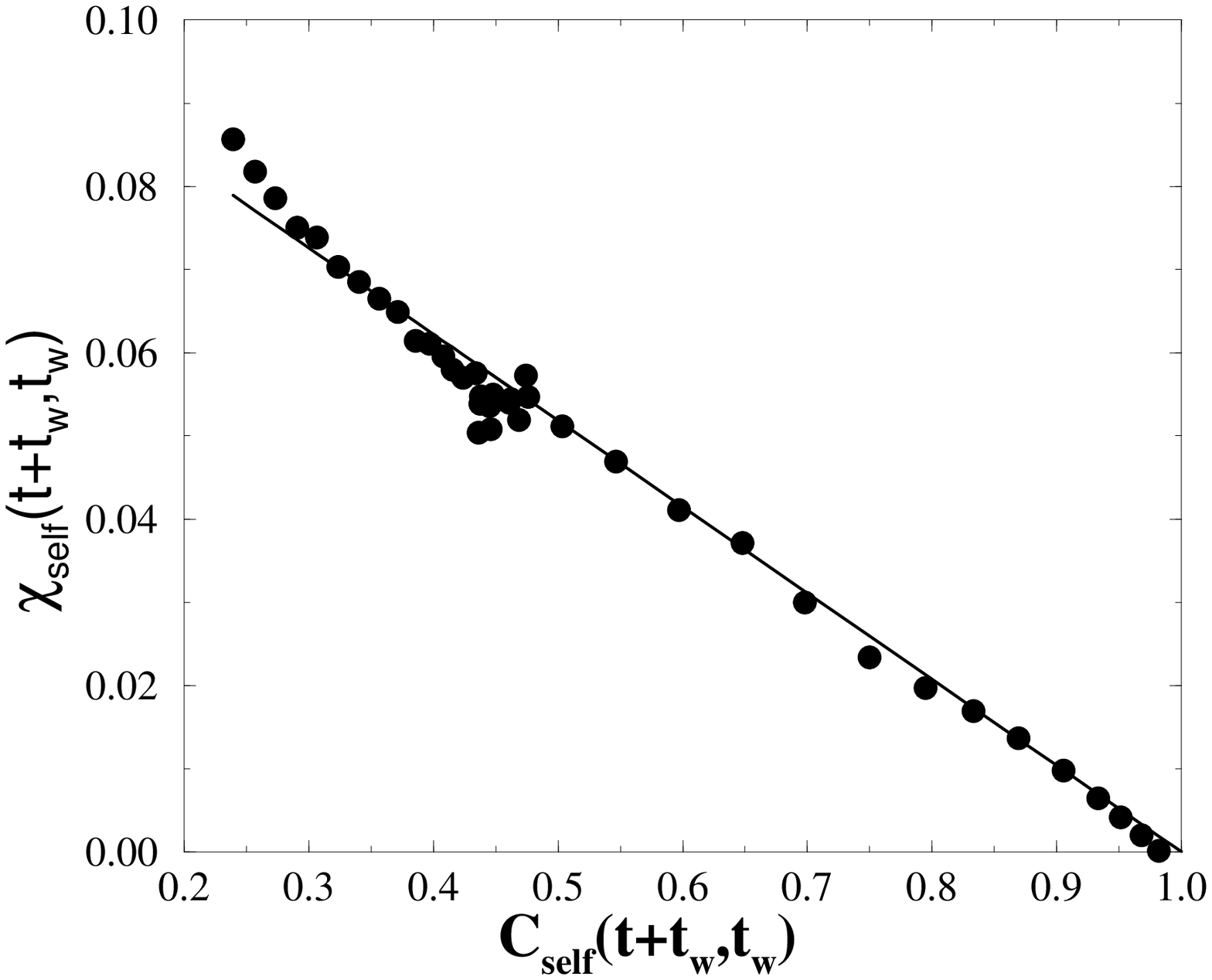,width=7cm,angle=-90}
    \hspace*{0.1cm}
    \psfig{figure=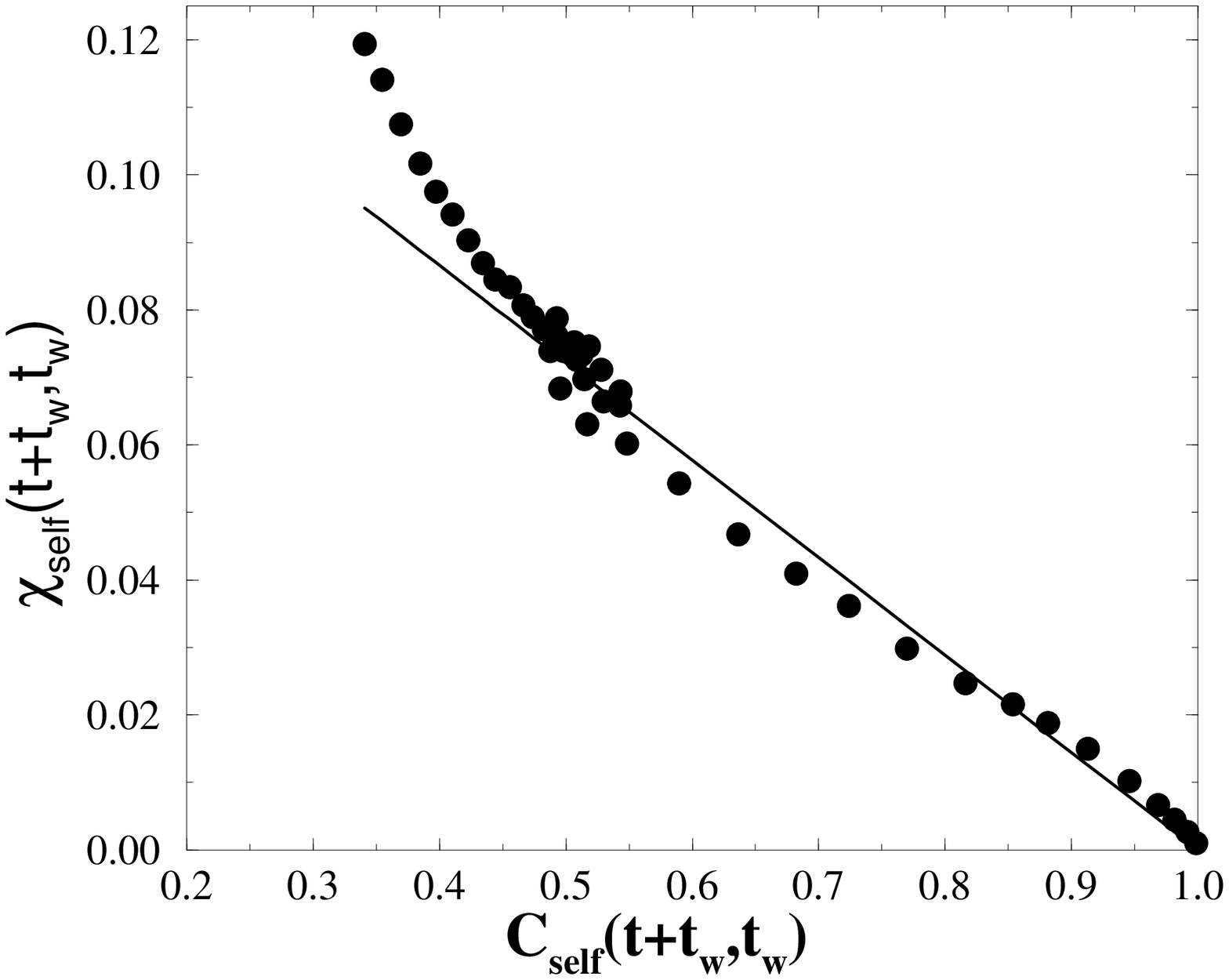,width=7cm,angle=-90}
}
\end{center}
\caption{
Parametric plots of the correlation and response functions for the
collective (upper panels) and the self (lower panels) variables, at
the two temperature $T_1=2900$~$K$ (left panels) and $T_2=2500$~$K$;
the slope $\frac{d\chi}{dC}$ is proportional to $-1/T_{eff}$.
Straight lines are the predictions of the equilibrium FDT theorem
where the slope is proportional to $-1/T$. At the lower temperature,
$T_{eff}<T$ at long times (low values of correlation).  }
\label{fig:Teff}
\end{figure}

The parametric plots of $\chi$ versus $R$ for the quenches at the
higher temperature $T_1$ are unusual (Fig.~\ref{fig:Teff}, left
panels), in the sense that although we know the system is aging, the
slope of the plots is consistent with a $T_{eff}=T$ along all the
curves as in equilibrium FDT. This behavior is similar to the results
of ref.~\cite{fdt_strong} for the Fredrickson-Andersen model, but
contrasts with our expectation of a $T_{eff}$ higher than the $T_1$.
The parametric plots of $\chi$ versus $R$ at $T_2$
(Fig.~\ref{fig:Teff}, right panels) yields even more astonishing
results: while at short times (high values of C) the slope
$\frac{d\chi}{dC}$ is equal to $-V_0\frac{1}{k_B T_2}$ yielding the
usual equilibrium FDT relation, in the long time region (low values of
$C$), the parametric plot bends upward, i.e. $T_{eff}$ is lower than
the bath temperature.

In conclusion, we have shown that for a realistic model of strong
liquid, the aging dynamics is not consistent with the commonly
observed 1RSB scenario.  Moreover, the behavior of the
fluctuation-dissipation ratio $X$ contrasts with the possibility of
interpreting $T_{eff}$ as a temperature. From the point of view of
landscape theory, this effect could be due to a profound and still to
be understood difference in the way strong and fragile landscapes are
explored.  As for strong liquids barriers to relaxation grow slower
than for fragile liquids while exploring deeper regions of the
landscape, the possibility remains in fact open that activated
processes are ``too fast'' and don't allow to describe the system in
terms of long-lived quasi-states~\cite{metabiroli} and to define
therefore a simple extension of landscape thermodynamics to OOE
conditions. The results reported in this letter clearly contradict the
general consensus built in recent years in the study of models of
fragile liquids~\cite{fdt_goodexamples,fsT,fdtRUOCCO}. This pushes for
a deeper theoretical understanding of the behavior of the strong
glass-forming liquids.

We acknowledge support from MURST PRIN 2000, INFM-PRA-HOP,
INFM-Initiative Parallel Computing.

\newpage

\newpage

\newpage


\begin{references}
\bibitem{note:2T}
This assertion should be limited to ``not so far'' from equilibrium
systems, as it is expected that two parameters are not enough to
describe in general OOE systems~\protect\cite{free_land}; moreover,
this is certainly not true for systems with infinite replica symmetry
breaking~\protect\cite{infbreak}
\bibitem{note:freexpl}
With saddle-dominated entropic dynamics we mean that the system is
exploring the landscape through accessible saddles that lead to lower
energy minima; this search is entropic, i.e. activation is due to the
probability of finding a direction toward lower energies. In this
regime, temperature controls just the velocity of the trajectory in the
configuration space and small differences in kinetic temperature are
irrelevant on logaritmic scale. If kinetic temperature is too low,
after a first phase of entropic search saddles are not accessible
anymore and the system has to cross barriers to relax to lower minima;
it is therefore expected a relevant dependence of the dynamics on the
kinetic temperature. Such scenario has been proposed for a fragile
glass-former in~\protect\cite{KST} and verified
in~\protect\cite{LAPRSS}.
\bibitem{note:linreg}
We have checked that the response functions scaled by $V_0$ still
collapse on each other for values of the field up to three times
bigger than the ones employed in the simulations.
\bibitem{note:stn}
The importance of a good signal-to-noise ratio stems from the
incredible difficulty in obtaining good statistics for the response
and the correlation functions. We have been forced to average over
$\sim 10^4$ quenches for each data set in order to obtain data of good
quality; this has meant one year of continuous running on a
proprietary cluster of six Compaq Alpha and on eight nodes of the
SGI-Origin at Cineca.
\bibitem{note:biggertw}
Choosing $t_w$ 10 times bigger would have requested simulations 10
times longer to observe the same amount of decorrelation (assuming
that structural decorrelation times scale as $t_w$). This would have
been unfeasible with present computer facilities. It could be achieved
at the expenses of decreasing by 10 the number of independent
realizations and hence increasing the noise level.
\end{references}
\end{document}